\documentclass[twocolumn]{article}

\pdfoutput=1
\nonstopmode{}

\usepackage{parskip}

\usepackage{algpseudocode}
\usepackage[utf8]{inputenc}
\usepackage{pgfplots}
\usepackage{caption}
\usepackage{amsmath}

\begin{document}

\title{Stream-based aggregation of unreliable heterogeneous network links}

\author{
    Michał Zieliński\\
    Jagiellonian University\\
    michal@zielinscy.org.pl \\
    \\
    Supervisor: \\
    Michał Staromiejski \\
    mistar@tcs.uj.edu.pl
}


\maketitle
\begin{abstract}
Last mile link is often a bottleneck for end user. However, users typically have multiple ways of accessing the Internet (cellular, ADSL, public Wifi). This observation led to creation of protocols like mTCP~\cite{mtcp} or R-MTP~\cite{rmtp}. Current bandwidth aggregation protocols are packet based.
However, this is not always practical --- for example, non-TCP protocols are often blocked on firewalls. Moreover, a lot of effort was devoted over the years into making single-path TCP work well over various types of links.
In this paper we introduce protocol which uses multiple TCP streams to establish single reliable connection attempting to maximize bandwidth and minimize latency.
\end{abstract}


\section{Introduction}
In this paper we introduce protocol which uses multiple TCP streams (with different last mile link) to establish single reliable connection attempting to maximize bandwidth and minimize latency. The goal of the protocol is to transmit data from one host to another using multiple network paths.

Usually one host is a mobile device (for example a laptop or a mobile wireless router). Another host is a proxy hosted on a server with fast connection to the Internet. The proxy forwards connections from mobile device to hosts on the Internet, which presumably do not support this protocol. In this scenario client opens multiple TCP or TLS over TCP connections (each on a different interface) to the proxy and transmits data via them.

When packet is to be transmitted it is queued in an internal waiting queue of bounded size. When link is ready to transmit data, it chooses packet based on \textit{packet scheduling algorithm}. It does not necessarily choose first packet in queue. For example there may be one link with small latency and one with bigger. When the link with bigger latency is ready to transmit, it still may be preferable to send first packet via the link with smaller latency (which may not be yet ready).

\section{Comparison to other protocols}

By using existing transport protocol (TCP), our protocol leverages decades of work dedicated
to improving TCP over various link types.
To illustrate the point, there are 14 different TCP congestion control algorithms in the Linux kernel alone.
As TCP-based web is probably the most important application of the Internet, we should expect futher development of TCP and similar protocols (such as QUIC).

In addition, in virtually all environments TCP port 443 is not filtered on firewalls, while
UDP or IP based protocols often are.

If queue and reorder buffers are large enough our protocol uses all available bandwidth
(as it always sends packets if there is any ready-to-send stream).
This is in constrast to existing network-level striping protocols, which must
deal with discrepancies in assumptions of higher level protocols or need to reimplement
congestion control themselves.

Tunneling multiple TCP over single TCP stream may cause head-of-line blocking~\cite{hol}.
One can be avoid this by using other transport protocols such as SCTP or QUIC.
This is particulary important as we think that QUIC may become dominant transport
protocol for HTTP/2.0 (it is already commonly used by Google Chrome when connecting to Google services).

Hsieh and Sivakumar~\cite{ptcp} argue that striping transmission over multiple TCP links is unoptimal.
However, use of EDPF in our protocol mitigates ``Data rate differential'' problem
raised in their work.
Futhermore, our experiments have shown that even when ``dumb striping'' is used this effect is not significant.


\section{Link characterization}

Let us consider amount of data received over a link $T$, assuming sending side continously transmits. It is an increasing function $\text{size}_T(t)$ of bytes over time. We define \textit{link characteristic function $\operatorname{char}_T$} as a derivative of $\text{size}_T$ over time.

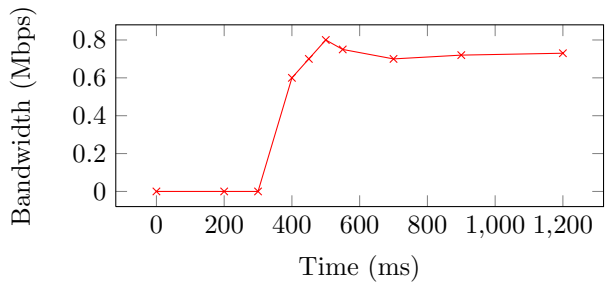
\begin{figure}
  \begin{tikzpicture}
	\begin{axis} [
		xlabel=Time (ms),
		ylabel=Bandwidth (Mbps), height=4cm, width=\columnwidth]
	\addplot[color=red,mark=x] coordinates {
	  (0,0)
	  (200,0)
      (300,0)
	  (400,0.6)
	  (450,0.7)
	  (500,0.8)
	  (550,0.75)
      (700,0.70)
      (900,0.72)
	  (1200,0.73)
	};
	\end{axis}
  \end{tikzpicture}

  \caption{Example characteristic function of real link}\label{fig:figreal}
\end{figure}

\begin{figure}
\begin{tikzpicture}
	\begin{axis} [
		xlabel=Time (ms),
		ylabel=Bandwidth (Mbps), height=4cm, width=\columnwidth]
	\addplot[color=red,mark=x] coordinates {
	  (0,0)
	  (200,0)
      (399,0)
	  (400,0.70)
	  (800,0.70)
	  (1200,0.70)
	};
	\end{axis}
\end{tikzpicture}
\caption{Simple link characteristic function}\label{fig:figsimple}
\end{figure}
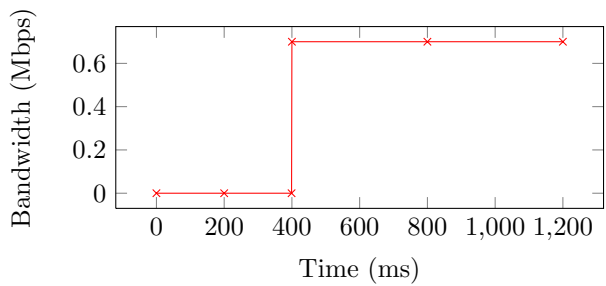

For purpose of scheduling algorithms, we assume that in an instant of time link has constant latency and bandwidth. This leads to a simple link characteristic function (figure~\ref{fig:figsimple}).

\begin{figure}
\begin{tikzpicture}
	\begin{axis} [
		xlabel=Time (ms),
		ylabel=Bandwidth (Mbps), height=4cm, width=\columnwidth]
	\addplot[color=red,mark=x] coordinates {
	  (0,0)
	  (200,0)
      (399,0)
	  (400,0.70)
      (599,0.70)
	  (600,1.1)
	  (1200,1.1)
	};
	\end{axis}
\end{tikzpicture}
\caption{Characteristic function of aggregated simple links}\label{fig:figmulti}
\end{figure}
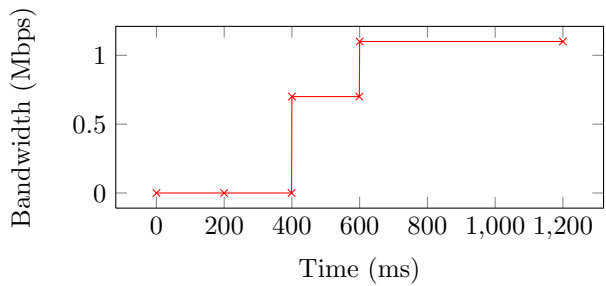

The characteristic function of an ideal link aggregation algorithm would be the sum of aggregated link characteristic functions (figure~\ref{fig:figmulti}).

\section{Packet scheduling}

Our scheduling algorithm is based on EDPF~\cite{edpf} (Earliest Delivery Path First) significantly modified for streaming transport protocols (e.g. TCP).

When a link $T_0$ becomes ready-to-send we need to choose which packet is to be send via it. To accomplish this goal the algorithm keeps some state:

\begin{enumerate}
\item a queue of packets along with the link which they are currently scheduled for
\item for each link $T_i$:
  \begin{enumerate}
  \item $E_{T_i}$ --- an estimated number of bytes which is currently travelling over $T_i$
  \item $\operatorname{char}_{T_i}$ --- an estimated \textit{link characteristic function} for $T_i$
  \end{enumerate}
\end{enumerate}

Packets in the queue are scheduled by the following algorithm:

\begin{algorithmic}
\State$E' \gets\mbox{ copy of }E$
\For{packet in the queue, starting from front}
   \State$T \gets \mbox{link with earliest estimated delivery}$
   \State\Comment{(based on $E'$ and $\operatorname{char}$)}
   \State$\mbox{assign }T\mbox{ to the current packet}$
   \State$\mbox{increase }E'_{T}\mbox{ by the size of the packet}$
\EndFor{}
\end{algorithmic}

Estimated number of bytes $E_{T_i}$ is increased when a packet is transmitted and decreased on every access by current estimated bandwidth multiplied by the time elapsed from the last access.

Estimated delivery time can be computed as

\begin{equation} \label{eq:delivery}
\Delta t_T(s + E_T) = \text{size}_T^{-1}(s + E_T)
\end{equation}

Data transmitted over a link $T$ is the integral of its characteristic function:

\begin{equation} \label{eq:transmit}
\text{size}_T(\Delta{} t) = \int^{\Delta t}_0 \operatorname{char}_{T} dt
\end{equation}

We assume that the link characteristic function is simple --- it has constant bandwidth and latency (figure~\ref{fig:figsimple}).
Then, based on equations \eqref{eq:delivery} and \eqref{eq:transmit}, the following holds:

\begin{equation} \label{eq:simple}
\Delta t_T(\text{size}) = \text{latency} + \text{size} / \text{bandwidth}
\end{equation}

In our case:

\begin{equation} \label{eq:simple1}
\Delta t_T = \text{latency} + (s + E_t) / \text{bandwidth}
\end{equation}

\subsection{Bandwidth estimation}

Determining available bandwidth between two hosts on the Internet is hard~\cite{bandwidth}.
However, we only need accurate bandwidth estimation when a lot (comprable to the size of congestion window) of data is sent. Arrival time of small data bursts is dominated by the link latency.

Bandwidth estimation is updated only when TCP buffers of the operating system are filled. In this case the bandwidth is simply a rate in which OS empties the buffers.

\section{Implementation}

Our packet scheduling algorithm can be effeciently implemented by exploiting equation~\eqref{eq:simple}:

\begin{algorithmic}
\State$E' \gets\mbox{ copy of }E$
\State$L \gets\mbox{ empty priority queue (heap)}$
\For{link $T$}
   \State$\mbox{push T to } L \mbox{ with priority }\Delta t_T(E'_T)$
\EndFor{}
\For{packet in the queue, starting from front}
   \State$T \gets \mbox{pop-min from } L$
   \State$\mbox{assign }T\mbox{ to the current packet}$
   \State$\mbox{increase }E'_{T}\mbox{ by the size of the packet}$
   \State$T \gets \mbox{push } T\mbox{ to } L \mbox{ with weight }\Delta t_T(E'_{T})$
\EndFor{}
\end{algorithmic}

\section{Retransmission}

While decision to dispatch packet to some link is the best possible based on information we have, network condition change may be unnoticable for a long time. For example, it is not possible to immediately detect upstream cable being unplugged from Wifi AP we are using. This means that we may send packet to a link which latency has significantly increased (even to infinity, when link is never available again).

However our goal is to guarantee reliable delivery. In order to achieve that goal we introduce Acknowledgement packets and packet sequence IDs. When packet is sent via link, monotonically increasing seqence ID is prepended to it. Acknowledgement packets contain seqence ID of last received packet and are sent periodically. When acknowledgement for an interval of packets is not received for some time, the packet is readded to the waiting queue.

\section{Tunneling of VPN}

Possibility of tunneling of existing protocols would greatly increase utility of link aggregation protocol. However, we cannot simply tunnel IP packets over reliable connection. This can lead to ``TCP over TCP meltdown''~\cite{tcpovertcp} that happens when TCP congestion control from two layer interfere badly.

To overcome this issue we can terminate TCP connections on both sides and send raw TCP stream over aggregated link in a manner similar to how SOCKS transporent proxies work. As TCP requires ordered and reliable delivery and proposed aggregation protocol provides only reliability, additional reorder buffer has to be implemented.

In most cases additional flow control has to be implemented, for example one similar to SSH-2 Connection Protocol~\cite{ssh-connection}.

It can be shown that the size of the reorder buffer has to be of order of retransmission timeout multiplied by the fastest link bandwidth.


\section{Experimental data}

We have have tested two internet connections --- LTE (cellular) and Ethernet-over-Coaxial (wired).
Target host was running in nearby datacenter with substantially larger available bandwidth.
Tests marked x10 took 20 seconds and were repeated 10 times. We have averaged data from all measurements.
Tests marked x1 took 60 seconds and was performed once.
Moving average filter was applied to bandwidth results with window size equal to 100 ms (blue line) and 1000 ms (green line).

\includegraphics{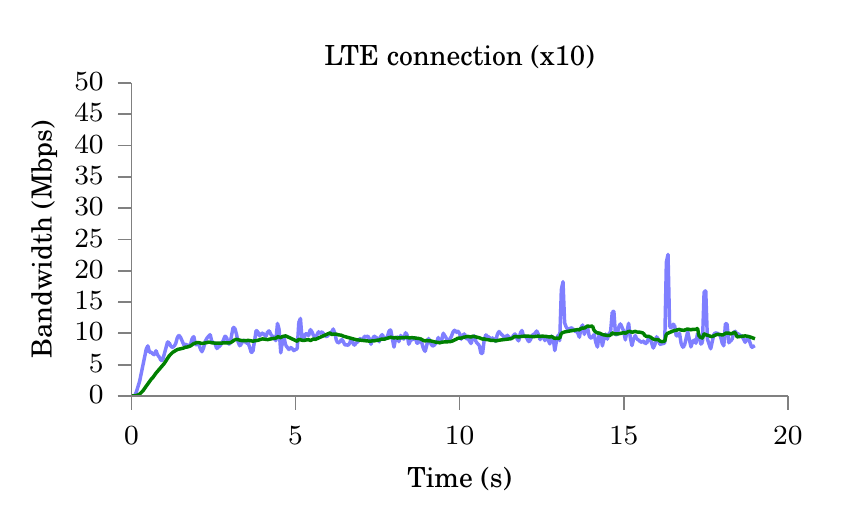}

\includegraphics{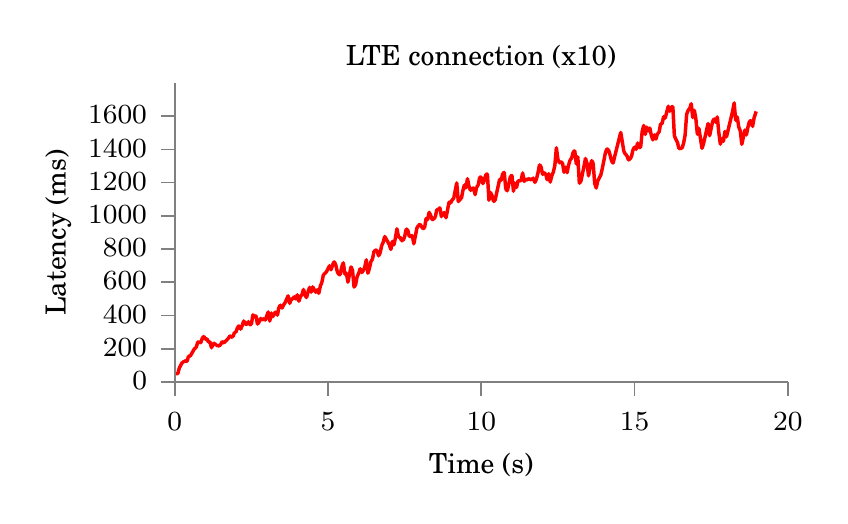}

LTE has fairly high latency and low bandwidth.

\includegraphics{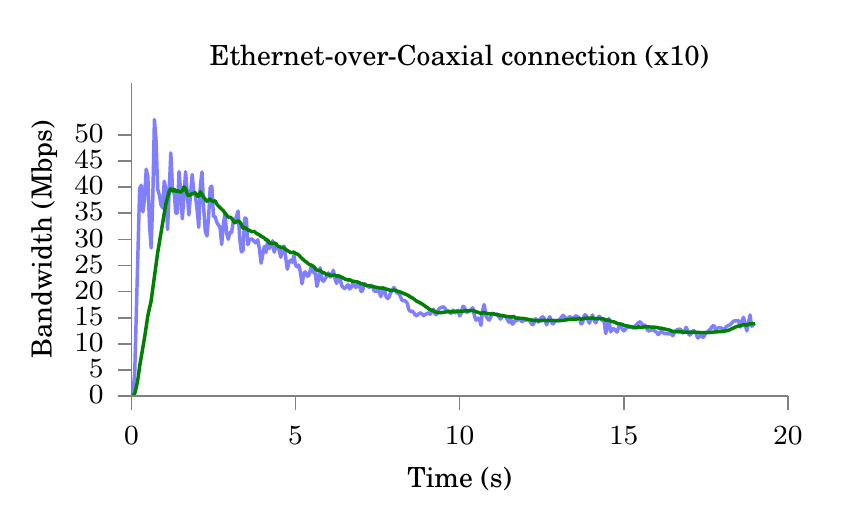}

\includegraphics{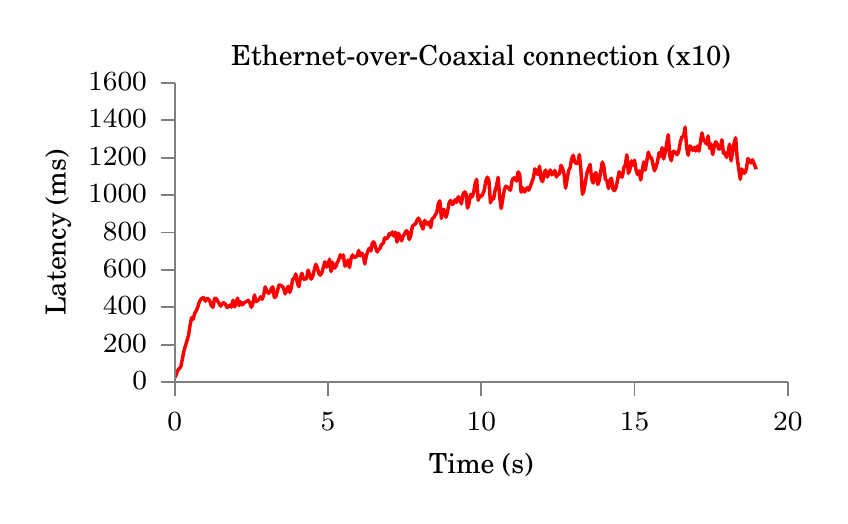}

EoC connection has smaller latency. However, large network buffers also increase latency during congestion.

\includegraphics{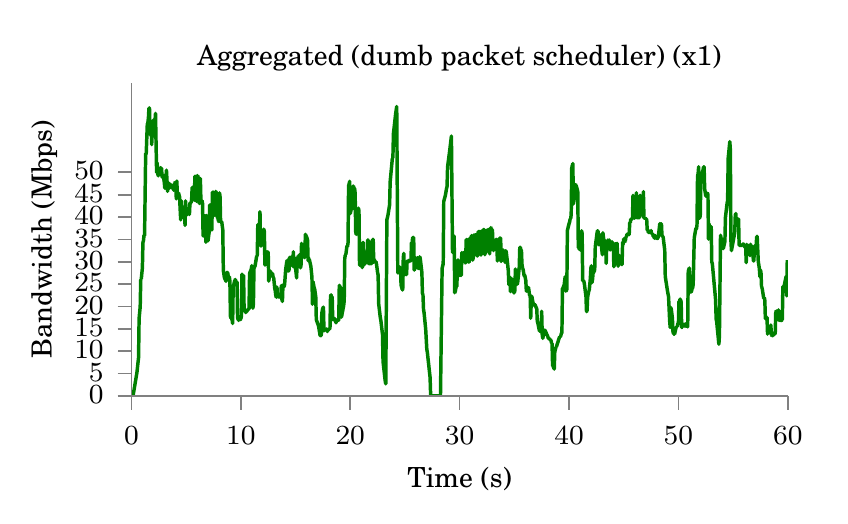}

\includegraphics{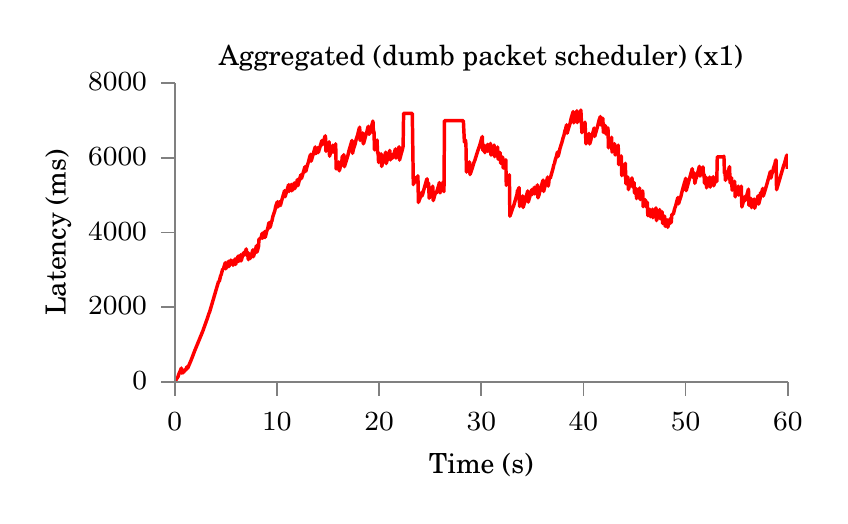}

\includegraphics{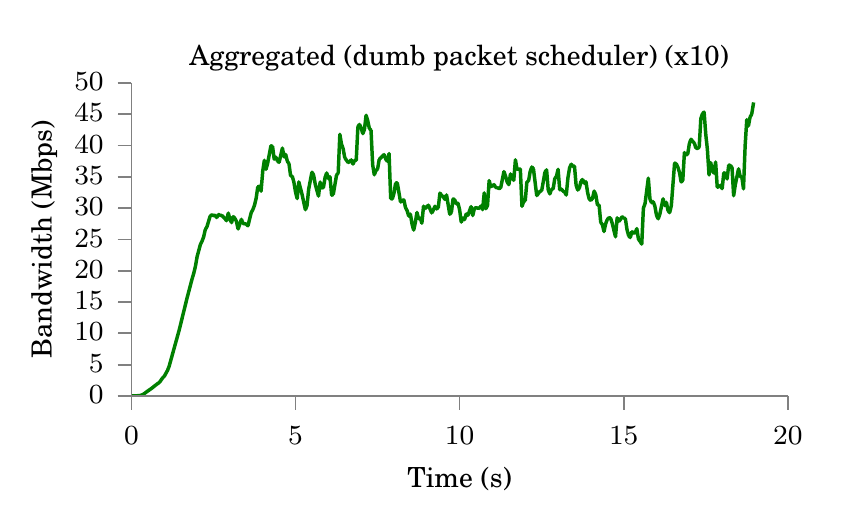}

\includegraphics{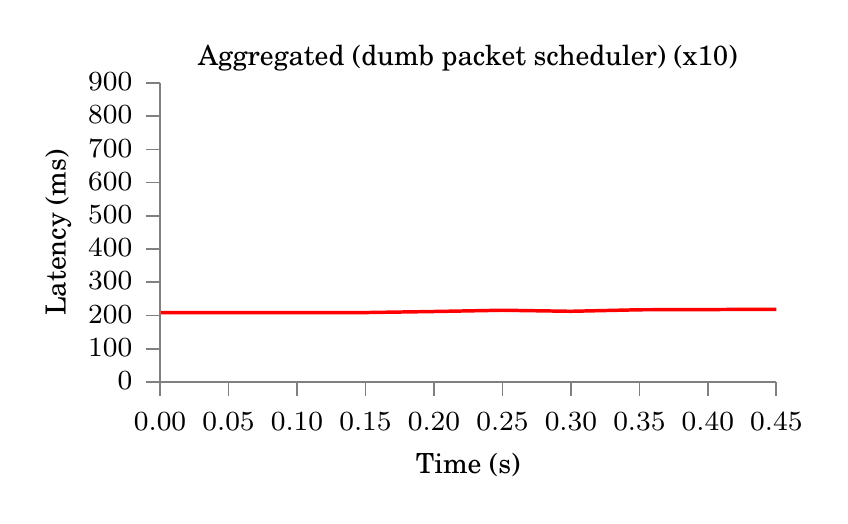}

\includegraphics{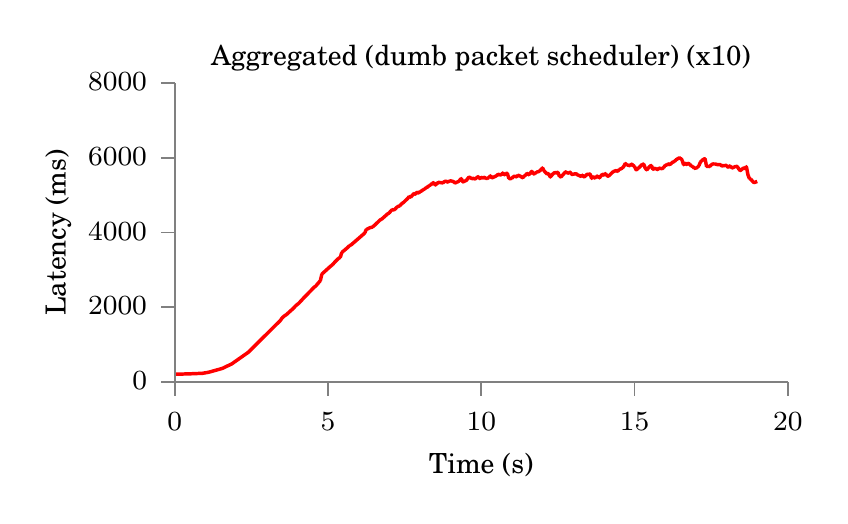}

We can see that dumb scheduling algorithm has latency worse than both links. Bandwidth jitter is also large.
Both effects can be caused by packets arriving out of order and waiting in the reorder buffer.
However, overall bandwidth is roughly equal to the sum of bandwidth of aggregated links.

\includegraphics{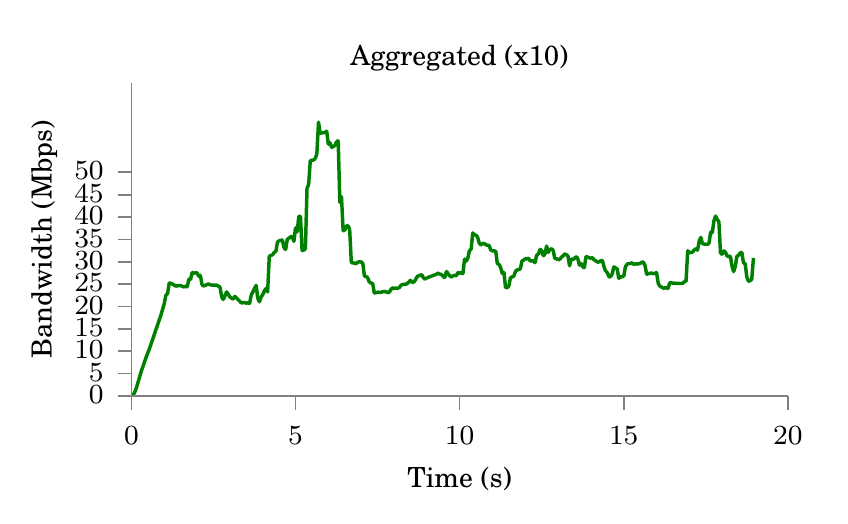}

\includegraphics{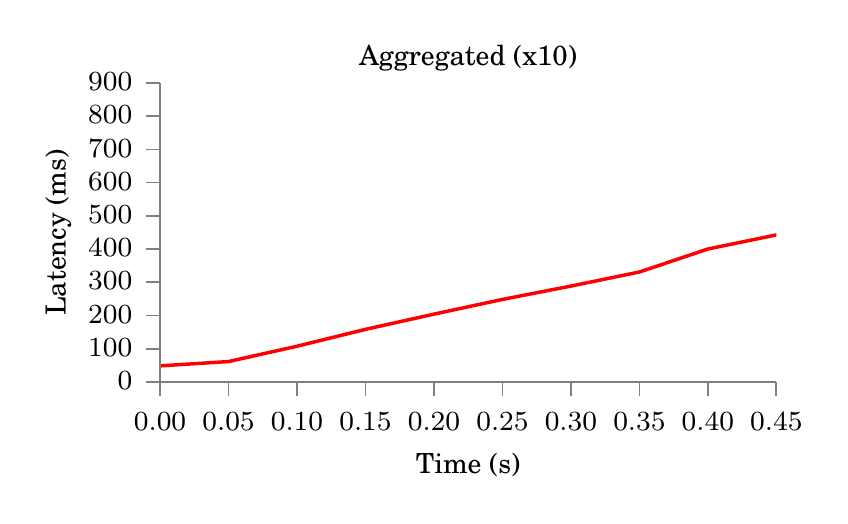}

\includegraphics{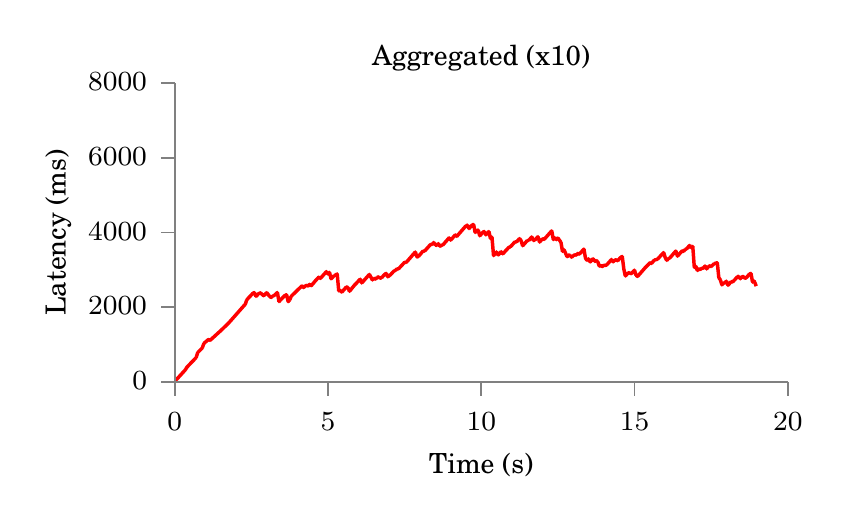}

Our packet scheduling algorithm also has bigger latency than both links during sustained transfer.
This is due to the packet scheduling being imperfect when network conditions are not constant.
However, latency of initial packets is equal to the latency of the link with better latency.

\section{Conclusion}

Our approach can achieve goals of more complex protocol in a simpler way by aggregating links above transport layer. Moreover, it has bigger potential for adoption in current networks.
However, due to effects like TCP over TCP meltdown\cite{tcpovertcp}, it is not well suited for tunneling of non-stream protocols such as UDP.\@

\end{document}